\documentclass[a4paper,11pt]{article}
\usepackage{subcaption}
\usepackage{epsfig, graphicx, subfloat, float}
\usepackage{latexsym}
\usepackage{cite}
\usepackage{amssymb}
\usepackage{latexsym}
\usepackage{mathrsfs}
\usepackage{amsmath}
\usepackage{color}
\usepackage[usenames]{xcolor}
\graphicspath{{Image/}}
\begin{document}
\title {Mass-Varying Dark Matter Induced Scalarization and Scalar Clouds around Black Holes}
\author{H. Mohseni Sadjadi\footnote{mohsenisad@ut.ac.ir}, M. Navid Gasemi Zad \footnote{meysam.navid@ut.ac.ir}
	\\ {\small Department of Physics, University of Tehran,}
	\\ {\small P. O. B. 14395-547, Tehran 14399-55961, Iran}}
\maketitle
\begin{abstract}
We consider a black hole solution embedded in a mass-varying dark matter (DM) halo and study
scalarization induced by DM. We derive the required conditions on the scalar–DM coupling
that allow for a regular bound-state scalar cloud. We find that such configurations exist only for quantized coupling values
in terms of the halo's intrinsic parameters (for Hernquist, this reduces to the inverse of the compactness).
This study provides a novel connection between DM phenomenology and scalarization around a black hole.
\end{abstract}

\section{Introduction}

The question of whether a stationary black hole can support regular, nontrivial scalar-field configurations
has a long history and is closely connected to the development of black-hole no-hair theorems.
Beginning with the pioneering works of Bekenstein \cite{bek1}, it was shown that asymptotically flat stationary black holes
cannot support canonical minimally coupled scalar fields which inherit the symmetries of the space-time and
whose self-interaction potentials satisfy suitable positivity conditions \cite{bek2}. Over the years, numerous models
have been proposed to test this conjecture and to find counterexamples by relaxing its core assumptions.
Modifications to the gravitational theory itself,
the introduction of non-canonical kinetic terms, or the inclusion of non-minimal couplings to curvature have
all been explored as possible avenues to evade the theorem. For a comprehensive review of such scenarios, see,
e.g. \cite{radu}, references therein, and \cite{circ1,circ2,circ3,cir4,cir5}.

Since galactic black holes are expected to reside inside dense dark matter (DM) distributions
\cite{Halo1,Halo2,Halo3,Halo4,Halo5,Halo6,Halo7,Halo8,Halo9,Halo10,Halo11,Halo12,Halo13},
it is interesting to investigate
whether interactions between DM and scalar degrees of freedom result in regular bound state scalar fields or
induce scalarization phenomena. Spontaneous scalarization may be induced by geometry, and matter fields
and environmentally triggered scalar instabilities \cite{scalar1,scalar2,scalar3,scalar4,scalar5,scalar6,scalar7,scalar8}.
In many of these scenarios, the scalar field acquires
a negative effective mass squared, producing tachyonic growth and eventually
leading to a static scalarized solution. DM halos surrounding astrophysical black holes provide a
natural environment in which such mechanisms may operate.

A static, asymptotically flat black hole solution in a spherically symmetric
DM halo may be derived by solving the Einstein equations by considering the DM energy momentum (EM) tensor
in the matter source sector \cite{Halo5,bh2,bh3,bh4}. Under these conditions, and for a canonical, minimally coupled massless scalar field, the original reasoning of Bekenstein remains applicable. Consequently, no regular, nontrivial, static scalar field configuration can be supported by such a black hole–DM system. The presence of the halo, although modifying the space-time geometry,
does not by itself open a channel for the existence of a regular scalar bound state in the minimal coupling case.

In this work, we want to investigate how this situation
changes once we allow a relation between the scalar field and DM.
Specifically, we consider a model of mass-varying DM (whose mass depends on the scalar field) \cite{mv1,mv2,mv3,mv4}.
The effect of this on the equation of motion of the scalar field is similar to a conformal coupling, which arises naturally
in interacting dark sector models and scalar-tensor cosmology, known as the screening mechanism \cite{khour1,khour2}. We study conditions required
for scalarization (emergence of a nontrivial scalar field in the black hole background) induced by this mass-varying DM.

The paper is organized as follows. In Sec. II, we present and solve the gravitational system
(in our study, we ignore the direct backreaction of the scalar field on the black hole metric) and discuss the black hole solution
and derive the scalar field equation in the context of mass-varying DM. We obtain the required conditions to circumvent the no-go theorem,
which forbids having a non-trivial regular scalar field. Conditions for
scalarization are studied. In Sec. III, we study the problem in the context of
a Sturm-Liouville eigenvalue problem and investigate the bound on the scalar-DM coupling parameter. By using the WKB approach,
we derive quantized scalar-DM couplings and also study asymptotic bound-state solutions. For the Hernquist model,
these results show that the scalar-dark matter coupling is quantized in terms of the inverse of the halo compactness.
Finally, in Sec. IV, we conclude our results.

We use units $\hbar=c=G=1$. Our convention for the metric signature is $(-,+,+,+)$.

\section{Mass-Varying DM and Field Equations}
We consider an asymptotically flat, spherically symmetric, static space-time
\begin{equation}\label{eq1}
ds^2=-f(r)e^{-2\delta(r)}dt^2+\frac{1}{f(r)}dr^2+r^2d\Omega^2,
\end{equation}
where $f(r)$ and $\delta(r)$ are analytical functions satisfying  $f(r\to \infty)=1$ and $\delta(r\to \infty)=0$.
We assume the existence of an event horizon at $r_h$, where $f(r_h)=0$ and $\delta(r_h)\in \mathbb{R}$.
The action is given by
\begin{equation}\label{eq2}
S=\int d^4x\sqrt{-g}\left(\frac{R}{16\pi}-\frac{1}{2}g^{\mu \nu}\partial_\mu\varphi\partial_\nu\varphi\right)+S_{DM},
\end{equation}
where $S_{DM}$ is the DM sector and $\varphi$ is a massless scalar field.
Einstein's equations, $G_{\mu \nu}=8 \pi T_{\mu \nu}$, yield
\begin{eqnarray}\label{eq3}
&&\frac{df(r)}{dr}=8\pi r T^t_t+\frac{1-f(r)}{r}\nonumber \\
&&\frac{d\delta(r)}{dr}=\frac{4\pi r}{f(r)}\left(T^t_t-T^r_r\right).
\end{eqnarray}
We assume that the EM tensor in the Einstein equations arises solely from
the DM halo, i.e., $T^{(\varphi)}_{\mu \nu}\ll T_{\mu \nu}$,
where $T^{(\varphi)}_{\mu \nu}$ and $T_{\mu \nu} $ are the EM tensors of the scalar field and DM, respectively.

The Bekenstein no-go theorem \cite{bek2} forbids regular, nontrivial scalar field configurations in static black hole backgrounds.
Here, we consider a scalar field in the context of mass-varying DM and investigate the conditions required to circumvent
this no-go theorem, allowing a regular bound state scalar field.

In mass-varying DM models, the DM mass $m(\varphi)$
depends on the scalar field $\varphi$. The scalar field EM tensor satisfies \cite{mv1,mv2,mv3}
\begin{equation}\label{eq4}
\nabla_\mu {T^{(\varphi)}}^\mu_\nu=-T A_{,\varphi}\varphi_{,\nu},
\end{equation}
where $T$ is the trace of DM EM tensor
$T=T^\mu_\mu$, and $A_{,\varphi}=\frac{d\ln m(\varphi)}{d\varphi}$.
A similar approach can be adopted by considering a metric conformal coupling in the DM sector
through which the scalar field couples to DM \cite{khour1}, leading to the same scalar field equation of motion (\ref{eq4}) \cite{sad}.
The massless scalar field equation is given by
\begin{equation}\label{eq5}
\Box \varphi=-T A_{,\varphi}.
\end{equation}
We assume that $A_{,\varphi}(\varphi=0)=0$, hence $\varphi=0$ is the trivial solution to (\ref{eq5}). We also assume
$-T A_{,\varphi\varphi}(\varphi=0)<0$, leading to a negative squared mass for $\delta\varphi$ fluctuations.
Integrating over a region $\mathcal{R}$, and using the Killing horizon properties, from (\ref{eq5})
we obtain \cite{kh}
\begin{equation}\label{eq6}
\int_{\partial \mathcal{R}} \varphi \nabla^\mu \varphi \sqrt{h} n_\mu d^3 \sigma-\int_\mathcal{R} \nabla_\mu \varphi \nabla^\mu \varphi
\sqrt{-g} d^4x= -\int_\mathcal{R} \varphi T A_{,\varphi} \sqrt{-g} d^4x.
\end{equation}
$\mathcal{R}$ is bounded by the black hole horizon and spatial infinity (and in time, by two partial Cauchy surfaces \cite{Hawk} whose contributions cancel each other \cite{radu}). $h_{\mu\nu}$
is the induced metric on the boundary. We have a Killing event horizon, and the scalar field is assumed to vanish
(or fall off sufficiently fast) at infinity. We obtain
\begin{equation}\label{eq7}
-\int_\mathcal{R} \nabla_\mu \varphi \nabla^\mu \varphi
\sqrt{-g} d^4x= -\int_\mathcal{R} \varphi T A_{,\varphi} \sqrt{-g} d^4x.
\end{equation}
In the absence of mass-varying DM, the right-hand side vanishes, forcing $\varphi=const.$.
A nontrivial solution for $\varphi$ therefore requires $\int_\mathcal{R} \varphi T A_{,\varphi} \sqrt{-g} d^4x>0$.

To solve (\ref{eq5}), we need an expression for $T$. From (\ref{eq3}), and the continuity equation $T^\mu_{r;\mu}\simeq 0$,
we derive the trace of the DM's EM tensor \cite{DN}
\begin{equation}\label{eq8}
T=\frac{1}{r^3}e^{\delta(r)}\left(e^{-\delta(r)}r^4T^r_r\right)'-\frac{3f(r)-1}{2f(r)}\left(T^r_r-T^t_t\right).
\end{equation}

In the following, we choose the DM particle mass as \cite{scalar1,sad}
\begin{equation}\label{mass1}
m_{\chi}(\varphi)=m_{\chi}(0)\exp\left(\frac{1}{2}\alpha \varphi^2\right),
\end{equation}
with $\alpha<0$, leading to the linear coupling
\begin{equation}\label{eq9}
A_{,\varphi}=\alpha \varphi.
\end{equation}
Note that $A_{,\varphi}(\varphi=0)=0$ and $A_{,\varphi\varphi}(\varphi=0)=\alpha<0$ are consistent with the required conditions for scalarization.
The DM becomes lighter
in regions where $\varphi^2$ is larger. In the limit $|\alpha| \varphi^2\ll1$ (\ref{mass1}) reduces to
\begin{equation}\label{mass2}
m_{\chi}(\varphi)\approx m_{\chi}(0)(1+\frac{1}{2}\alpha \varphi^2),
\end{equation}
which implies the interaction $\mathcal{L}_{int.}=-\frac{\alpha}{2} m_{\chi}(0)\varphi^2 \bar{\chi}{\chi}$ between
the scalar and fermionic DM field, and $\mathcal{L}_{int.}=-\frac{\alpha}{4} m_{\chi}(0)\varphi^2 {\chi^2}$ for the scalar DM.

We note that Eq. (\ref{mass1}) corresponds to the coupling function (\ref{eq9}) for which the scalar field equation (\ref{eq5}) becomes exactly linear in $\varphi$. This will allow us to formulate the scalar field as a linear bound-state problem and enables the analytic and numerical determination of the discrete coupling values supporting regular scalar configurations. More general coupling functions would generally lead to nonlinear scalar field equations. A general parametrization would be
\begin{equation}
m_\chi(\varphi)= m_\chi(0)F(\varphi),
\end{equation}
with $F(0)=1$. For $\varphi=0$ to be a solution we need $F'(0)=0$ and to have a tachyonic state, we require $\alpha\equiv \frac{d^2\ln F(\varphi)}{d\varphi^2}|_{\varphi=0}<0$. For sufficiently small $\varphi$, one then has: $\ln m_\chi(\varphi)=\ln m_\chi(0)+\frac{1}{2}\alpha \varphi^2+\mathcal{O}(\varphi^3)$, which implies $A_{,\varphi}=\alpha \varphi + \mathcal{O}(\varphi^2)$.
Substituting this into the scalar field equation (\ref{eq5}), which is the basis of our calculations, yields, to linear order in
$\varphi$, precisely the same scalar field equation obtained from the exponential coupling in (\ref{mass1}). Away from the linearized regime, however, higher-order terms in $F(\varphi)$ generally render the scalar field equation nonlinear, and the resulting finite-amplitude scalar configurations depend on the specific form of the coupling function.

In the following, we also adopt the $\delta(r)=0$ solution, which implies $T^r_r=T^t_t$. Hence (\ref{eq8}) simplifies to
\begin{equation}\label{eq10}
T=-r\rho'(r)-4\rho(r),
\end{equation}
where $\rho(r)$ is the energy density. Thus, for $\delta(r)=0$, only the energy density is needed to compute the trace of the EM tensor.
For example, for the Hernquist DM profile \cite{Halo1}
\begin{equation}\label{eq11}
-T^t_t=\rho(r)=\rho_s\left( \frac{r}{r_s}\right)^{-1}\left(1+\frac{r}{r_s}\right)^{-3},
\end{equation}
where $\rho_s$ is a real number with dimensions of density, and $r_s$ is a length scale, we find
\begin{equation}\label{eq12}
T=-\frac{3M_s r_s^2}{2\pi r^5\left(1+\frac{r_s}{r}\right)^4},
\end{equation}
where we have defined
\begin{equation}\label{eq13}
M_s:=4\pi\int_0^ \infty \rho r^2 dr=2\pi \rho_s r_s^3.
\end{equation}
An exact black hole solution to (\ref{eq3}) in DM (\ref{eq11}) is given by
(note that we have a class of solutions $h(r)=f(r)+\frac{\beta}{r}$, where $\beta$ is a constant)
\begin{equation}\label{eq14}
f(r)=1-\frac{2M_0}{r}-\frac{2M_sr}{(r+r_s)^2},
\end{equation}
where $M_0$ is a real constant with dimension of mass (or equivalently length in our units).
Note that $M_s$ is not the mass inside the proper volume, but it can be interpreted as the ADM mass
of the DM. In our solution, the ADM mass in this asymptotically flat space-time is
\begin{equation}\label{eq15}
M_{ADM}=\lim_{r\to \infty} \frac{r}{2}\left(1-f(r)\right)=M_0+M_s.
\end{equation}
Thus $M_s$ is the mass contribution from the halo, and $M_0$ is the bare black hole mass.
We assume $M_0\ll r_s$ and also $M_s\ll r_s$. The event horizon, $r_h$,
is the solution of
\begin{equation}\label{h}
1-\frac{2M_0}{r}-\frac{2M_sr}{(r+s)^2}=0.
\end{equation}
Hence for the solution (\ref{eq14}), we obtain
\begin{equation}\label{eq33}
r_h\simeq 2M_0\left(1+\frac{4M_0M_s}{r_s^2}\right)\simeq 2M_0.
\end{equation}

We assume that the scalar field inherits the space-time symmetries, such that $\varphi=\varphi(r)$, and therefore the mass
for the DM particle is $m_\chi(r)=m_\chi(\varphi(r))$.
The scalar field (which can also be regarded as the $l=0$ (s) mode of a general static field) satisfies
\begin{eqnarray}\label{eq16}
\frac{d}{dr}\left(r^2f(r) \frac{d\varphi(r)}{dr}\right)&=&-\alpha r^2 T \varphi(r)\nonumber \\
&=&\alpha r^2 \left(r\rho'(r)+4\rho(r)\right)\varphi(r),
\end{eqnarray}
 where the term $-\alpha T$ acts as a coordinate-dependent effective mass squared term for the scalar field, and is negative
 for $ \alpha T>0$. In general, there is no analytic solution to (\ref{eq16}), but one can approximate solutions in different regions.
For $\alpha=0$, as mentioned in the literature \cite{radu}, the near-horizon non-trivial solution also exhibits a logarithmic singularity:
\begin{equation}\label{eq17}
\varphi(r)\propto \frac{1}{r_h^2 f'(r_h)}\ln(z-1),
\end{equation}
where $z:=\frac{r}{r_h}$, implying that the system does not support a non-trivial regular scalar field. However, for $\alpha \neq 0$,
one can use the Frobenius ansatz $\varphi(r)=(z-1)^s\sum_{n=0}^\infty a_n(z-1)^n$,
which yields $s^2=0$, indicating the existence of a regular branch:
\begin{equation}\label{eq18}
\varphi(r)=a_0\left[ 1-\frac{\alpha r_h T(r_h)}{f'(r_h)}(z-1)+O((z-1)^2)\right],
\end{equation}
where $a_0$ is a constant. For the Hernquist profile (\ref{eq11}), (\ref{eq18}) reduces to
\begin{equation}\label{eq19}
\varphi(r)=a_0\left[ 1+\frac{3\alpha \mathcal{C} r_h}{2\pi r_s}(z-1)+O((z-1)^2)\right],
\end{equation}
where we have defined
\begin{equation}\label{eq27}
\mathcal{C}:=\frac{M_s}{r_s},
\end{equation}
which is dubbed the halo compactness and quantifies how concentrated the halo is. An upper bound for the compactness
$\mathcal{C} \leq0.092 $ is reported in \cite{range1}, with the Event Horizon Telescope (EHT) observations. Also, the galactic dynamics bounds
$\mathcal{C} \geq 10^{-4}$ \cite{Halo2}. Note that for $\alpha=0$ the regular solution is only a trivial one $\varphi(r)=a_0$.

Introducing $\psi=r \varphi(r)$ and the tortoise coordinate $\frac{dr*}{dr}=\frac{1}{f(r)}$, we can convert (\ref{eq16}) to a Schrodinger like equation
\begin{eqnarray}\label{eq20}
\frac{d^2\psi}{d{r*}^2}&=&-\alpha Tf(r)\psi+\frac{f(r)f'(r)}{r}\psi\nonumber \\
&=& V_{eff.}\psi,
\end{eqnarray}
where
\begin{equation}\label{eq21}
V_{eff.}:=-\alpha Tf(r)+\frac{f(r)f'(r)}{r}.
\end{equation}
For large $|\alpha|$ and $T<0$, the effective potential develops a potential well. Here, this is due to scalar-DM coupling,
but in the scalar-tensor model, this can arise from nonminimal coupling to the geometry. For example, in the scalar Gauss-Bonnet model, where a term
$-\frac{1}{2}\alpha \mathcal{G} \varphi^2$ is included in the Lagrangian (with  $\mathcal{G}$  the Gauss-Bonnet invariant), the term
$-\alpha T$ in (\ref{eq21}) is replaced by $\alpha \mathcal{G}$ \cite{scalar6}. For a Schwarzschild black hole with mass $M_0$,
$\alpha \mathcal{G}=48 \frac{\alpha M_0^2}{r^6}$.

\begin{figure}[H]
\centering
\includegraphics[width=0.9\textwidth]{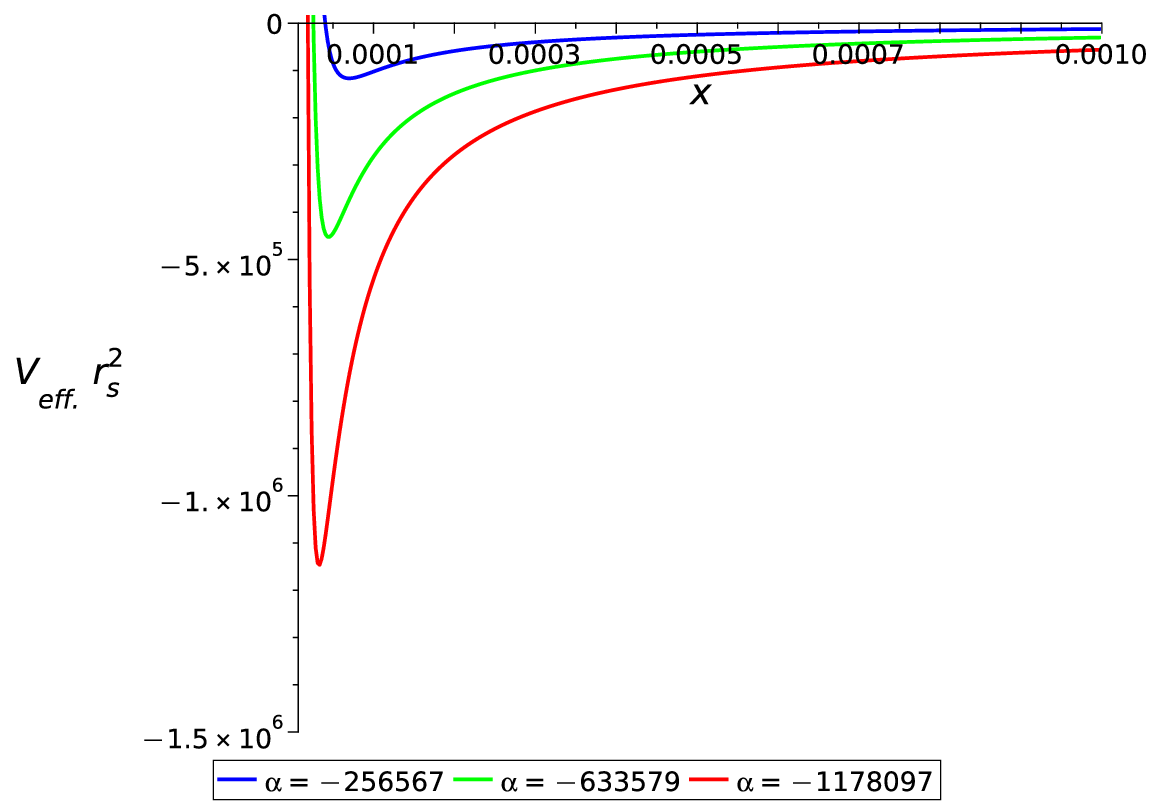}
\caption{Effective potential as a function of $x=\frac{r}{r_s}$ for three values of $\alpha$}
\label{fig1}
\end{figure}

\begin{figure}[H]
\centering
\includegraphics[width=0.9\textwidth]{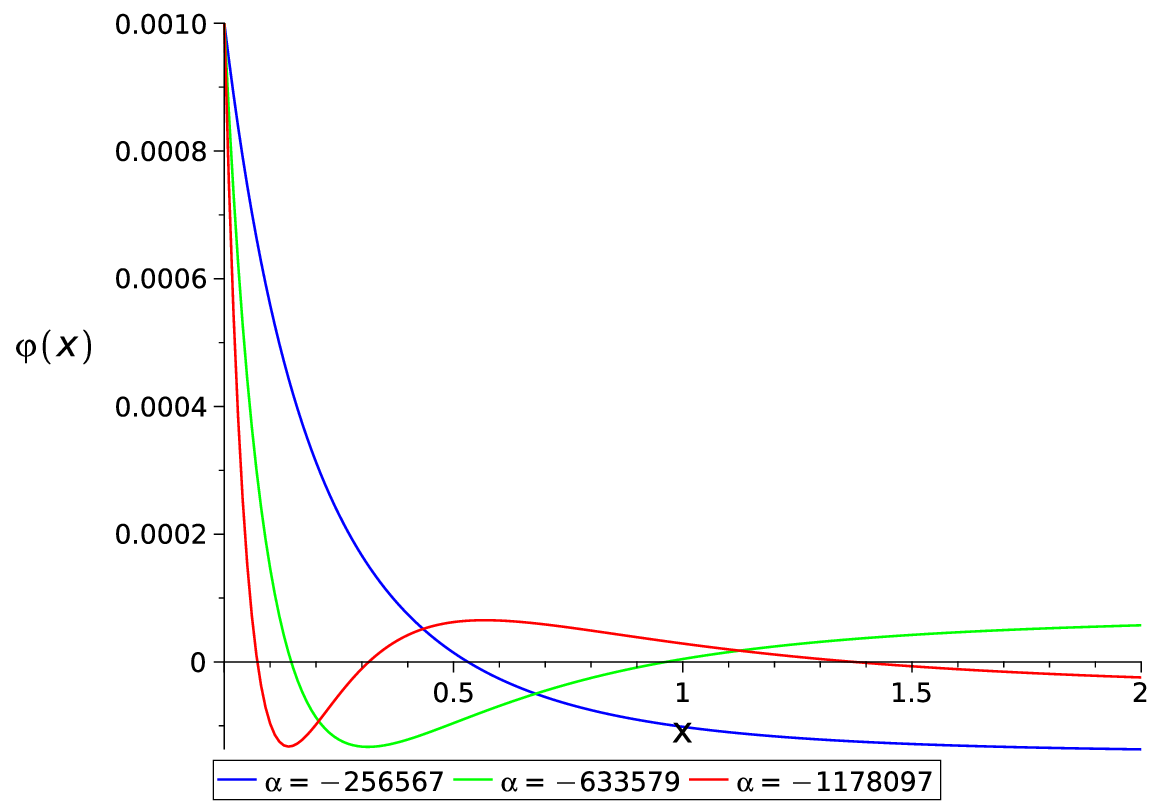}
\caption{Scalar field as a function of $x=\frac{r}{r_s}$ for three values of $\alpha$.}
\label{fig2}
\end{figure}

The behavior of the potential for the Hernquist profile (\ref{eq11}) and
for three negative values of $\alpha$ is depicted in Fig. (\ref{fig1}), where we have set
$\mathcal{C}=10^{-4}$ and $\frac{M_0}{r_s}=10^{-8}$.
The larger $|\alpha|$, the deeper the potential well, which facilitates the existence of bound-state scalar fields.
In Fig. (\ref{fig2}) we illustrate the behavior of the
scalar field using the same parameters as in Fig. (\ref{fig1}) with initial conditions $\varphi(x=10^{-6})=0.001$ and
$\frac{d\varphi(x)}{dx}(x=10^{-6})=1$, showing that larger $|\alpha|$ leads to more nodes.
We cannot find analytic solutions to the equations (\ref{eq16}) or (\ref{eq21}) across the entire region outside the black hole.
To gain some insight into the parameter $\alpha$, we proceed as follows.

\section{Scalar-DM Coupling}
From (\ref{eq16}) we derive
\begin{equation}\label{eq22}
\int_{r_h}^\infty \varphi(r) \frac{1}{r^2}\frac{d}{dr}\left(r^2f(r)\frac{d\varphi}{dr}\right)r^2{dr}=
-\alpha\int_{r_h}^\infty \varphi^2(r) T r^2 dr,
\end{equation}
which, for a regular scalar field at the horizon and decaying at infinity, yields
\begin{equation}\label{eq23}
\alpha=\frac{\int_{r_h}^\infty f(r)(\varphi'(r))^2 r^2dr}{\int_{r_h}^\infty \varphi^2(r)T r^2dr}.
\end{equation}

\begin{figure}[H]
\centering
\includegraphics[width=0.9\textwidth,height=0.4\textheight,keepaspectratio]{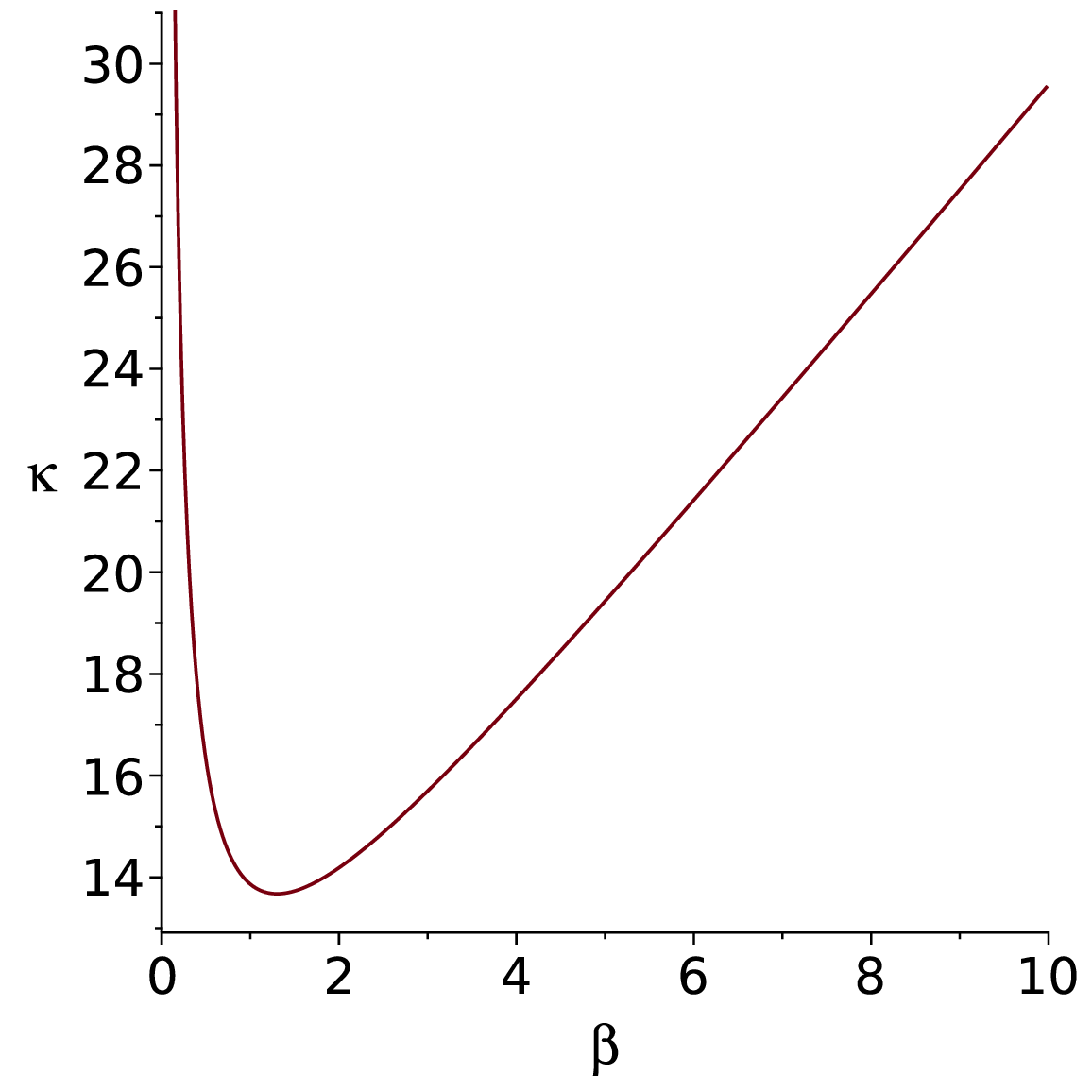}
\caption{$\kappa$ as a function of $\beta$}
\label{fig3}
\end{figure}

If $\alpha=0$, then $\int_{r_h}^\infty f(r)(\varphi'(r))^2 r^2dr=0$, which implies $\varphi'(r)=0$. Thus, without scalar-DM
interaction, only the trivial solution $\varphi=0$ is allowed. For a nontrivial solution to exist and for $T<0$ (as in the case of (\ref{eq12})),
$\alpha$ must be negative: $\alpha<0$. The operator
\begin{equation}\label{eq24}
\mathcal{L}=\frac{1}{r^2T}\frac{d}{dr}\left(r^2f(r)\frac{d}{dr}\right)
\end{equation}
is self-adjoint with respect to the inner product $<\phi_1|\phi_2>=-\int_ {r_h}^\infty \phi_1(r)\phi_2(r)r^2T dr$. Considering
the Rayleigh–Ritz variational method,
we derive the following inequality:
\begin{equation}\label{eq25}
|\alpha_0|\leq \frac{\int_{r_h}^\infty f(r) (\Psi'(r))^2 r^2 dr}{-\int_{r_h}^\infty \Psi^2(r) T r^2dr },
\end{equation}
where we have assumed $T<0$. $\Psi(r)$
is any trial function satisfying the boundary conditions, i.e., regular at the horizon and decaying at infinity.
For a black hole in the  Hernquist DM, and using the trial function $\Psi(r)=e^{-\beta\frac{r}{r_s}}$ with $\beta>0$, along with the
 assumptions $M_s\ll r_s$ and $M_0\ll r_s$, we obtain
\begin{equation}\label{eq26}
|\alpha_0|\leq \frac{\int_{2M_0}^\infty \frac{1}{r_s^2} \left(1-\frac{2M_0}{r}-\frac{2M_s r}{(r+r_s)^2}\right)e^{-\frac{2\beta r}{r_s}}dr}
{\int_{2M_0}^\infty \frac{3M_s r_s^2}{2\pi r^3\left(1+\frac{r_s}{r}\right)^4}e^{-\frac{2\beta r}{r_s}}dr}\simeq \kappa \mathcal{C}^{-1}.
\end{equation}
In terms of the exponential integral function, $\mathrm{Ei}_{a}(x)=\int_{1}^{\infty} e^{-xt}t^{-a}dt$,
$\kappa$ is given by
\begin{equation}\label{eq28}
\kappa:=2.55 \beta  \left(\beta^{2} {\mathrm e}^{ 2 \beta} \left(\beta + 1.5\right)
\mathrm{Ei}_{ 1}\! \left( 2 \beta \right)+ 0.125- 0.5 \beta^{2}- 0.5 \beta \right).
\end{equation}
The function $\kappa(\beta)$ is plotted in Fig. (\ref{fig3}).
It has a minimum value $\kappa_{min.}=\kappa(\beta\simeq 1.25)\simeq 13.67$.  Therefore, we find $|\alpha_0|<13.67 \mathcal{C}^{-1}$.

\subsection{$\alpha$ Quantization}

One may use equation (\ref{eq20}) via the WKB method
to determine the allowed values of $\alpha$ that yield bound-state scalar field solutions \cite{scalar6}. In particular, a standard
 WKB approximation for the bound-state field configurations of the effective potential $V_{eff.}(r)$ gives the quantization condition
\begin{equation}\label{eq29}
\int_{r*_{-}}^{r*_{+}}\sqrt{-V_{eff.}}dr*=(n+1-\tilde{\delta})\pi,
\end{equation}
where $r*_{\pm}$ are turning points, $n$ is a non-negative integer, and $0<\tilde{\delta}<1$. In the eikonal regime (where the coupling term dominates over the other potential terms throughout most of the WKB integration range),
and assuming $T(r\to \infty)=0$, and
recalling that $f(r_h)=0$, (\ref{eq29}) can be approximated as \cite{scalar6}
\begin{equation}\label{eq30}
\int_{r_h}^\infty \sqrt{\frac{\alpha T}{f(r)}}dr  =(n+\frac{3}{4})\pi
\end{equation}
(see also Fig.(\ref{fig1})). Here we have set $\tilde{\delta}=1/4$, although $\tilde{\delta}=3/8$ is also reported to be more consistent with a
numerical analysis in a scalar-Gauss-Bonnet model \cite{scalar6}.
For the Hernquist density, substituting (\ref{eq12}) and (\ref{eq14}) into (\ref{eq30}) yields
\begin{equation}\label{eq31}
\sqrt{-\frac{3\alpha M_s}{2\pi r_s^3}} I=(n+\frac{3}{4})\pi,
\end{equation}
where
\begin{equation}\label{eq32}
I=\int_{r_h}^\infty G(r,M_s)dr=\int_{r_h}^\infty \frac{1}{\sqrt{\left(1-\frac{2M_0}{r}
-\frac{2M_sr}{(r+r_s)^2}\right){\frac{r}{r_s}(\frac{r}{r_s}+1)^4}}}dr,
\end{equation}
and $r_h$ is the horizon radius given by (\ref{eq33}).
The above integral does not have a closed-form analytical solution. However, assuming  $M_s\ll r_s$ and $M_0\ll r_s$,
we may approximate it as $I=I^{(0)}+I^{(1)}M_s+\mathcal{O}(M_s^2)$,
where
\begin{equation}\label{eq34}
I^{(0)}:=\int_{2M_0}^\infty \frac{1}{\sqrt{\left(1-\frac{2M_0}{r}\right){\frac{r}{r_s}(\frac{r}{r_s}+1)^4}}}dr.
\end{equation}
Using the change of variable $t=\frac{r-2M_0}{r_s+2M_0}$, we find the solution in terms of the beta function $B$:
\begin{eqnarray}\label{eq35}
I^{(0)}&=&r_s\left(1+\frac{2M_0}{r_s}\right)^{-\frac{3}{2}}\int_0^\infty t^{-\frac{1}{2}}(1+t)^{-2}dt \nonumber\\
&=&r_s\left(1+\frac{2M_0}{r_s}\right)^{-\frac{3}{2}}B(\frac{1}{2},\frac{3}{2})\simeq \frac{\pi}{2}r_s-\frac{3\pi M_0}{2}.
\end{eqnarray}
From $I^{(1)}=\frac{dI}{dM_s}|_{M_s=0}$, we obtain
\begin{equation}\label{eq36}
I^{(1)}=\int_{2M_0}^{\infty}\frac{\partial G(r,M_s)}{\partial M_s}|_{M_s=0} dr-\frac{\partial r_h}{\partial M_s}G(2M_0,0),
\end{equation}
which, after some computation, reduces to
\begin{equation}\label{eq37}
I^{(1)}=r_s^{\frac{5}{2}}\left[\int_{2M_0}^{\infty}\frac{r^2}{(r+r_s)^4(r-2M_0)^{\frac{3}{2}}}dr -
 \lim_{r\to 2M_0}\frac{8M_0^2}{(r_s+2M_0)^4\sqrt{r-2M_0}}\right].
\end{equation}
Defining $u=r-2M_0$ and $A=2M_0+r_s$, (\ref{eq37}) becomes:
\begin{equation}\label{eq38}
I^{(1)}=r_s^{\frac{5}{2}}\int_0^ \infty \frac{1}{u^{\frac{3}{2}}}\left[\frac{(u+2M_0)^2}{(u+A)^4}-\frac{4M_0^2}{A^4}\right].
\end{equation}
Note that the singular term is cancelled out by the singular part of the first integral.
Using the identity $\int_0^{\infty} u^{\alpha-1}(u+A)^{-4}du=A^{\alpha-4}B(\alpha,4-\alpha)$, we find
\begin{equation}\label{eq39}
I^{(1)}=\frac{\pi r_s^{\frac{5}{2}}}{16(2M_0+r_s)^{\frac{9}{2}}}\left[r_s^2+24M_0r_s-96M_0^2\right].
\end{equation}
Putting everything together, in the limit $M_0\ll r_s$ and $M_s\ll r_s$, we obtain
\begin{equation}\label{eq40}
I=\frac{\pi r_s}{2}-\frac{3\pi M_0}{2}+\frac{\pi M_s}{16}+\mathcal{O}\left(\frac{M_0M_s}{r_s}\right)\simeq \frac{\pi r_s}{2}.
\end{equation}
Therefore, from (\ref{eq31}) we obtain
\begin{equation}\label{eq41}
\alpha_n=-\frac{8\pi}{3} (n+\frac{3}{4})^2\frac{1}{\mathcal{C}}.
\end{equation}
Thus, we conclude that the discrete eigenvalues
$\alpha_n$ are inversely proportional to the halo compactness. For $n\gg 1$, where our approximation method is more accurate,
we have
\begin{equation}
 \alpha_n\approx-\frac{8\pi}{3} n^2\frac{1}{\mathcal{C}}.
 \end{equation}
Note that to obtain (\ref{eq41}), we have used the simplified integral, which assumes that the effective potential is dominated by the $\alpha $ term, while the other terms are subleading in the eikonal limit.

To examine whether this dependence of $\alpha$ solely on the inverse of the halo compactness, to leading order in $r_s$, is a characteristic of the Hernquist model, we briefly investigate a more general profile. For the Dehnen density, given by \cite{deh}
\begin{equation}\label{c1}
\rho=\frac{(3-\gamma)M_sr_s}{4\pi r^\gamma(r+r_s)^{4-\gamma}},
\end{equation}
(\ref{eq31}) becomes
\begin{equation}\label{c2}
\sqrt{\frac{\alpha(3-\gamma)(\gamma-4)M_s}{4\pi r_s^3}}I= (n+\frac{3}{4})\pi.
\end{equation}
Assuming that the metric and the horizon in the integral may be approximated by those of the bare black hole, we have
\begin{eqnarray}\label{c3}
I=I^{(0)}&=& \int_{2M_0}^\infty\frac{dr}{\sqrt{(1-\frac{2M_0}{r})(\frac{r}{r_s})^\gamma(1+\frac{r}{r_s})^{5-\gamma}}}\nonumber \\
&=&\frac{\pi}{2}r_s^{\frac{5}{2}}(r_s+2M_0)^{\frac{\gamma}{2}-2}(2M_0)^{\frac{1-\gamma}{2}} {}_{1}F_{2}(\frac{\gamma-1}{2},\frac{1}{2};2;-
\frac{r_s}{2M_0}),
\end{eqnarray}
where ${}_{1}F_{2}$ is the Gauss hypergeometric function. Since ${}_{1}F_{2}(0,\frac{1}{2};2;-
\frac{r_s}{2M_0})=1$ for $\gamma=1$, we recover the result obtained for the Hernquist model (\ref{eq35}):
$I^{(0)}=\frac{\pi}{2}r_s\left(1+\frac{2M_0}{r_s}\right)\simeq \frac{\pi}{2}r_s$.
Now let us consider another value of $\gamma$, for example $\gamma=2$. After some computation,
using ${}_{1}F_{2}(\frac{\gamma-1}{2},\frac{1}{2};2;-
\frac{r_s}{2M_0})\sim \frac{2}{\pi \sqrt{\frac{r_s}{2M_0}}}\ln \left(\frac{2r_s}{M_0}\right)$, which is valid for $\frac{r_s}{2M_0}\gg 1$,
we obtain
\begin{equation}\label{c4}
\alpha_n=- 2\pi^3(n+ \frac{3}{4})^2 \frac{1}{\mathcal{C}\ln^2 (\frac{2r_s}{M_0})}.
\end{equation}
This agrees with a dimensional analysis from which one finds that $I$, in (\ref{c3}), is generally of the form
$I= F\left( \frac{r_s}{M_0}\right)r_s$, where $F$ is a function. Substituting this into (\ref{c2}), we arrive
at $\alpha_n\propto \frac{(n+\frac{3}{4})^2} {F^2\left(\frac{r_s}{M_0}\right)\mathcal{C}}$.
For the Hernquist model, $F\left( \frac{r_s}{M_0}\right)\simeq \frac{\pi}{2}$.

\subsection{Asymptotic Solution}
For the Hernquist density profile (\ref{eq12}), the scalar field equation is given by
\begin{equation}\label{eq42}
\frac{1}{r^2}\frac{d}{dr}\left(r^2\left(1-\frac{2M_0}{r}-\frac{2M_s r}{(r+r_s)^2}\right)\frac{d\varphi}{dr}\right)=
\alpha \frac{3M_s r_s^2}{2\pi r^5\left(1+\frac{r_s}{r}\right)^4}\varphi(r).
\end{equation}
For $r\gg 2M_0$, we introduce the change of variable $x=\frac{r}{r_s}$ and define $\mu^2=-\frac{3\alpha M_s}{2\pi r_s}$. The equation then reduces to
\begin{equation}\label{eq43}
x^2\frac{d^2\varphi}{dx^2}+2x\frac{d\varphi}{dx}=-\mu^2\frac{x}{(1+x)^4}\varphi.
\end{equation}
A solution which remains well-behaved for $x\ll1$ is
\begin{equation}\label{eq44}
\varphi(x)= c\frac{1+x}{x} W_M(\frac{1}{2}\mu,\frac{1}{2}, \frac{2\mu x}{1+x}),
\end{equation}
where $W_M$ is the Whittaker M function and $c$ is a real constant. In the limit $x\gg1$, we obtain
\begin{equation}\label{eq45}
\varphi(x\gg 1)=cW_M(\frac{1}{2}\mu,\frac{1}{2}, 2\mu)=2c \mu e^{-\mu}F(1-\frac{\mu}{2},2,2\mu),
\end{equation}
where $F$ denotes the confluent hypergeometric function. Therefore, to satisfy $\lim_{r\to \infty}\varphi(r)=0$, we must impose
$W_M(\frac{1}{2}\mu,\frac{1}{2}, 2\mu)=0$.
This condition yields discrete values of $\alpha$, which cannot be expressed in closed analytic form. However, for $\mu\gg 1$, we may employ the asymptotic approximation of the Whittaker function in (\ref{eq45}) to obtain the following equation
\cite{DLMF}
\begin{equation}\label{eq46}
\begin{split}
\varphi(x) &= c\frac{1+x}{x}\left(\frac{\mu}{2}\right)^{-\frac{1}{2}}
\sqrt{\frac{2\mu x}{1+x}}\,\Gamma(2)\\
&\quad\times\Bigg(J_1\Bigg(2\sqrt{\frac{\mu^2 x}{1+x}}\Bigg)
+ \mathrm{env}J_1\Bigg(2\sqrt{\frac{\mu^2 x}{1+x}}\Bigg)
O(\mu^{-\frac{1}{2}})\Bigg),
\end{split}
\end{equation}
where $J_1$ is the Bessel function of the first kind and  $env$ denotes its envelope. For $x\gg 1$ we find
\begin{equation}\label{eq47}
\varphi=2cJ_1(2\mu)+O(\mu^{-\frac{1}{2}}).
\end{equation}
Consequently, imposing $\varphi(r\to \infty)=0$ in the large $\alpha$ limit, and using the asymptotic form
$J_1(y\gg1)\sim \sqrt{\frac{2}{\pi y}}\cos(y -\frac{3\pi}{4})$ yields
\begin{equation}\label{eq48}
2\mu=2\sqrt{-\frac{3\alpha M_s}{2\pi r_s}}=(n+\frac{5}{4})\pi\simeq n \pi.
\end{equation}
Note that the condition $\mu\gg 1$ implies $n\gg 1$. Hence, in the large $n$ limit, we have
$\alpha_n \approx -\frac{\pi^3}{6}n^2 \frac{1}{\mathcal{C}}$.
This result differs from the one obtained via the WKB approximation, namely $\alpha_n \approx -\frac{8\pi}{3} n^2\frac{1}{\mathcal{C}}$,
by a numerical factor of order unity. In both approximations, $\alpha_n \propto -\frac{n^2}{\mathcal{C}}$.

\section{Conclusion and Outlook}

We have shown that a black hole solution in a DM halo can support regular static scalar field configurations in the context of mass-varying DM. The coupling between the scalar field and DM generates a negative effective mass-squared term
(a necessary condition for spontaneous scalarization), thereby allowing the system to evade the standard no-go theorems,
permitting regular bound-state scalar fields.

The scalar field equation was formulated as a Sturm–Liouville eigenvalue problem for the scalar-DM coupling parameter.
Using a variational analysis, we derived an upper bound on the magnitude of the fundamental eigenvalue.
For Hernquist DM density, the numerical value of this bound was obtained in terms of the compactness.
To determine the allowed couplings supporting bound-state scalar configurations, we analyzed the associated
Schrodinger-like equation. The WKB approximation yields a discrete spectrum, showing that the allowed couplings are quantized
in terms of the inverse of the halo compactness for the Hernquist profile. An independent analysis based on the asymptotic
solution of the scalar field leads to a similar scaling relation. For the more general Dehnen density profile, we showed that
the quantized coupling values are proportional not only to the inverse of the halo compactness but also to a combination of the DM halo characteristic radius and the bare black hole mass.

In the present work, the direct contribution of the scalar field in the EM tensor to the Einstein equations has been neglected. The scalar clouds obtained here can therefore be regarded as leading-order configurations on the black hole-DM background. An interesting extension would be to incorporate the gravitational backreaction of the scalar field perturbatively and determine the corresponding correction to the space-time metric: $g_{\mu \nu}=g^0_{\mu \nu}+\delta g_{\mu \nu}$, where the leading-order metric correction is sourced by the EM tensor of the scalar cloud. In particular, solving the Einstein equations to first order would give $\delta G_{\mu \nu}=  8\pi T^{(\varphi)}_{\mu \nu}[\varphi_0]$, where $\varphi_0$ is the scalar cloud found in the present work.
Such an analysis would make it possible to investigate possible direct gravitational signatures of the scalar cloud, including corrections to the photon-sphere and shadow structure and strong-field gravitational lensing \cite{Boz}, as well as modifications of the quasinormal-mode spectrum \cite{kok}, and the associated gravitational-wave signatures \cite{dreyer}. Since the scalar configurations occur only for the discrete coupling values obtained in this work, these observables could in principle provide a direct probe of the scalar-cloud configurations and their quantized couplings.

Finally, we note that the mass-varying DM model may influence the possible DM annihilation cross section, which depends
on the DM particle mass. The s-wave annihilation cross-section scales as
$\langle \sigma v \rangle=\langle \sigma v \rangle_0 (\frac{m_\chi (\varphi)}{m_{\chi}(0)})^{-2}= \langle \sigma v \rangle_0 e^{-\alpha \varphi^2}$
(under the assumption that $\langle \sigma v \rangle\propto m_{\chi}^{-2}$, which is typical for effective contact interactions or heavy mediator scenarios).
Therefore, for $\alpha<0$ which is required for scalarization,
the scalar field enhances the annihilation cross section (for example, in a galactic center with a black hole \cite{exe})
relative to the intrinsic cross-section $\langle \sigma v \rangle_0$.

A quantitative analysis of these observational implications provides a promising direction for further studies.

\end{document}